# Orbital Hall effect-driven spin-orbit torque enhancement in Ti-based systems via rare-earth interface engineering

Rohiteswar Mondal[1], Chennoju Raghu[1], Animesh Baral[1], Arabinda Haldar[2], Chandrasekhar Murapaka[1,*]

[1]Department of Materials Science and Metallurgical Engineering, Indian Institute of Technology Hyderabad, Kandi-502284, Telangana, India

[2]Department of Physics, Indian Institute of Technology Hyderabad, Kandi-502284, Telangana, India

*Corresponding author: mchandrasekhar@msme.iith.ac.in

## ABSTRACT

Harnessing the maximum potential of orbital currents in next-generation spin-orbitronic devices hinges critically on the efficiency of orbital-to-spin conversion. Despite possessing intrinsically large orbital Hall conductivities, light metals such as Ti translate this potential into spin-orbit torque (SOT) efficiency rather poorly, motivating substantial interest in engineering the interconversion process through interfacial design. Here, we investigate both orbital-to-spin and spin-to-orbital interconversion in Ti/Co heterostructures by inserting a Gd spacer and varying its thickness systematically. Using ferromagnetic-resonance-based spin and orbital pumping together with spin-torque ferromagnetic resonance (ST-FMR) measurements, we experimentally demonstrate both conversion pathways within the same trilayer architecture. We show that the large spin-orbit coupling associated with the 5*d* electronic states of Gd facilitates efficient orbital-to-spin interconversion, yielding a fivefold enhancement of the SOT efficiency relative to a Gd/Co reference. With Ti established as a critical orbital current source, comparison of the Ti/Gd/Co trilayer with Gd/Co and Ti/Co bilayer control samples reveals that efficient harnessing of orbital current transport requires not only a spacer of strong spin-orbit coupling but also smooth, well-ordered interfaces that preserve *d*-electron-mediated angular momentum transfer. These results identify the essential design criteria for a platform for orbital-to-spin conversion establishing rare-earth interlayer engineering as a viable strategy in next-generation orbitronic devices.

## INTRODUCTION

Current-induced spin-orbit torques (SOTs), generated via the spin Hall effect (SHE) in heavy 5*d* metals, underpin electrical control of magnetization switching and spin-wave excitation [1–5]. Although the large spin-orbit coupling (SOC) in Pt, Ta, and W yields sizable spin Hall conductivities (SHC, expressed as $\sigma_{SH}$) [2,6,7], it also imposes fundamental limitations on device performance. Elliott-Yafet spin relaxation limits the spin diffusion length in Pt or W [8,9], destructive interference among Berry-curvature hotspots suppresses the net spin Hall conductivity below its theoretical maximum [10], and interfacial spin-memory loss dissipates an additional 30-60% of injected spin current before it acts on the magnetization [11]. Moreover, the metastability and high resistivity of high spin Hall angle (SHA) $\beta$-W and the deep-level contamination risk of Pt in standard Si fabrication lines further constrain scalable CMOS integration [7]. These intrinsic limitations motivate a paradigm shift from spin current-based angular momentum transfer to orbital current-based angular momentum transfer. In this framework, orbital angular momentum (OAM) of conduction electrons, rather than spin, serves as the fundamental means of transporting transverse angular momentum. Analogous to the spin Hall effect (SHE) in heavy metals, the orbital Hall effect (OHE), driven by momentum-space orbital texture without any SOC requirement, has been identified as the microscopic origin of generating the orbital current in transition metals [10,12,13]. First-principles calculations show that certain 3*d* metals with relatively low SOC exhibit extraordinarily high orbital Hall conductivities (OHC) in the order of $10^4$ (ħ/e)(Ω·cm)$^{-1}$, one to two orders of magnitude larger than their spin counterparts, directly overcoming the Elliott-Yafet ceiling and Berry-curvature cancellation that inhibit degrade heavy metals [14,15]. These developments establish orbitronics [14,16], which utilizes orbital currents rather than spin currents, as a promising route toward next-generation, ultralow-power spin-orbitronic devices. The potential recently validated by a > 60% reduction in SOT-MRAM switching energy by using OHE in Ru and orbital Rashba effect at metallic interfaces [17–19].

Orbital torque (OT) operates through a three-step process: the OHE first generates an orbital current in the nonmagnetic (NM) layer, and this orbital current is then injected into the ferromagnet (FM), where its SOC converts OAM into spin angular momentum that couples to the magnetization [20,21]. This FM-SOC dependence, a fingerprint absent in conventional SHE-SOT, was confirmed in Ni/Ta and Cr/FM bilayers, where torque sign and magnitude follow the *L*-*S* correlation strength of the FM [22,23]. Experimental realization of OHE has since been demonstrated across a growing family of light 3$d$/4$d$ transition metals. Lyalin et al., detected orbital accumulation in Cr with an orbital diffusion length ($\lambda_L^{Cr}$) of ~6.6 nm via magneto-optical detection technique [24]. Orbital torque originating from OHE was confirmed in Zr by Fukunaga et al., and Yang et al., demonstrated that it can drive perpendicular magnetization switching at relatively lower current densities as compared to W [25,26]. Liu et al., gave an experimental proof of long-range orbital torque in Co/Nb bilayers [27]. These works collectively demonstrate that efficient orbital current generation is a generic property of weakly spin-orbit-coupled metals. Among these, titanium (Ti) is the benchmark: $\sigma_{\mathrm{OH}} \approx 3800\ (\hbar/e)(\Omega\cdot\mathrm{cm})^{-1}$ against $\sigma_{\mathrm{SH}} \approx -40\ (\hbar/e)(\Omega\cdot\mathrm{cm})^{-1}$, a ~100:1 OHC/SHC ratio [28]. Recently, orbital accumulation directly observed in Ti via MOKE and $\lambda_L^{Ti}$ of ~74 nm [28,29] was reported, which is one order higher magnitude beyond the spin diffusion length of any heavy metal. Interface engineering further facilitates smooth transfer of OAM: inserting a Pt conversion layer yields a multifold enhancement of effective orbital Hall angle, in similar range with topological insulators [30]. Moreover, a recent study has demonstrated a gigantic enhancement in magnetoresistance arising from a novel interaction between orbital magnets and orbital currents [31]. Recent experiments on surface-oxidized Cu/FM heterostructures suggest that OAM plays an active role in spin-orbit torque generation, with the underlying orbital current attributed to an interfacial orbital Rashba state [32]. Despite the large OHC predicted for light elements such as Ti, translating this into practical torque efficiency remains largely unexplored. First, the OHE cannot be directly harnessed through exchange coupling between OAM and local magnetization. Second, constructive addition of spin and orbital torques is required ($\langle L\cdot S\rangle_{\mathrm{NM}} \cdot \langle L\cdot S\rangle_{\mathrm{FM}} > 0$) to achieve high effective torque efficiency, thereby constraining material choice. In addition, orbital-angular-momentum carriers are more susceptible to interfacial disorder than spin carriers, resulting in significant transmission losses [23,33,34]. These intrinsic constraints collectively prevent orbital torques from realizing their theoretically predicated giant potential. Material systems that can simultaneously mitigate symmetry mismatch and interfacial scattering loss remain limited.

In this work, we address the above-mentioned bottlenecks by introducing a system that effectively improves orbital current transmission and orbital-to-spin conversion. We employed Ti as a dedicated orbital current source and demonstrated that a low-SOC ferromagnet, e.g. Co, is sufficient to sustain a sizable orbital torque without incorporating a heavy material of high SOC in the stack. To improve orbital-to-spin conversion efficiency, we introduce a rare-earth interlayer (IL), Gd, with relatively strong SOC facilitated by 5*d*-states which promotes orbital-to-spin current conversion. By systematically varying the interlayer thickness, we qualitatively map the efficiency and establish clear material-design rules. Our thickness-dependent measurements of Ti confirm that the dominant torque originates from the bulk OHE in Ti. Beyond establishing Ti as a robust orbital current source, quantitative demonstrations of high efficiency reveal rare-earth materials with high SOC *d*-states as a tunable knob for orbital-to-spin conversion efficiency, a design principle broadly transferable across the emerging family of light-metal orbitronic systems.

**Experimental procedure:**

The thin film heterostructures studied here are deposited on Si (naturally oxidized) substrates at room temperature using a DC magnetron sputtering system. The deposition conditions were optimized independently for each material to ensure reproducible film quality. We have prepared 5 sets of samples for this study. Sample set A consists of a Co (referred to as A1) reference sample, used to define the baseline for transverse voltage drop, and Ti(20)/Co(10) (referred to as A2), used to establish the contribution from Ti. To disentangle the inverse OHE (IOHE) contribution from that of the inverse SHE (ISHE), we have prepared sample set B with configuration Ti(20)/Gd($t_{Gd}$)/Co(10), and sample set C with configuration Gd($t_{Gd}$)/Co(10) (*t* in parentheses refers to the thickness variation). To probe bulk orbital contribution and qualitatively estimate the orbital diffusion length in Ti, we have deposited the sample set D, stacked as Ti($t_{Ti}$)/Gd(4)/Co(10). Finally, to support our results quantitatively, we microfabricated sample set E, with Ti(t)/Gd(4)/Co and Ti(20)/Co(10) configuration. All samples were capped with a Ti(3) layer to prevent oxidation, where the numerical values in parentheses denote the film thickness in nanometers. Spin-orbit pumping-driven electrical signals were characterized using a broadband ferromagnetic resonance (FMR) detection technique. A schematic of the measurement setup with the sample configuration is shown in Fig. 1 (a). The sample was mounted in a flip-chip geometry on a ground-signal-ground (GSG) coplanar waveguide (CPW), with the magnetic layer facing the signal line to maximize the *rf*-field coupling. The CPW was driven by an *rf* generator, and the transmitted signal was fed into a lock-in amplifier that locked the reference frequency of the external field modulation using a Helmholtz coil. A nanovoltmeter was used to measure the dc spin-orbit pumping voltage across the sample via IOHE/ISHE. The effective spin-orbit torque efficiency was quantified by spin-torque ferromagnetic resonance (ST-FMR). For these measurements, the heterostructure films were patterned into bar-shaped microdevices of dimensions 10×40 μm$^2$ using a maskless lithography system. Electrical contacts were established via GSG-configured pico-probes interfaced to a CPW, ensuring well-defined *rf* current injection and minimizing parasitic impedance contributions. ST-FMR spectra were recorded over the frequency range of 6-12 GHz under an in-plane magnetic field sweeping from 0-1200 Oe, applied at 45° to the current direction. For all samples, the resonance signal exhibited the highest signal-to-noise ratio at 9 GHz; therefore, we selected 9 GHz for the quantitative extraction of the effective torque efficiency ($\xi_{SOT}$) using the symmetric-to-antisymmetric line-shape decomposition method.

**Results and discussion:**

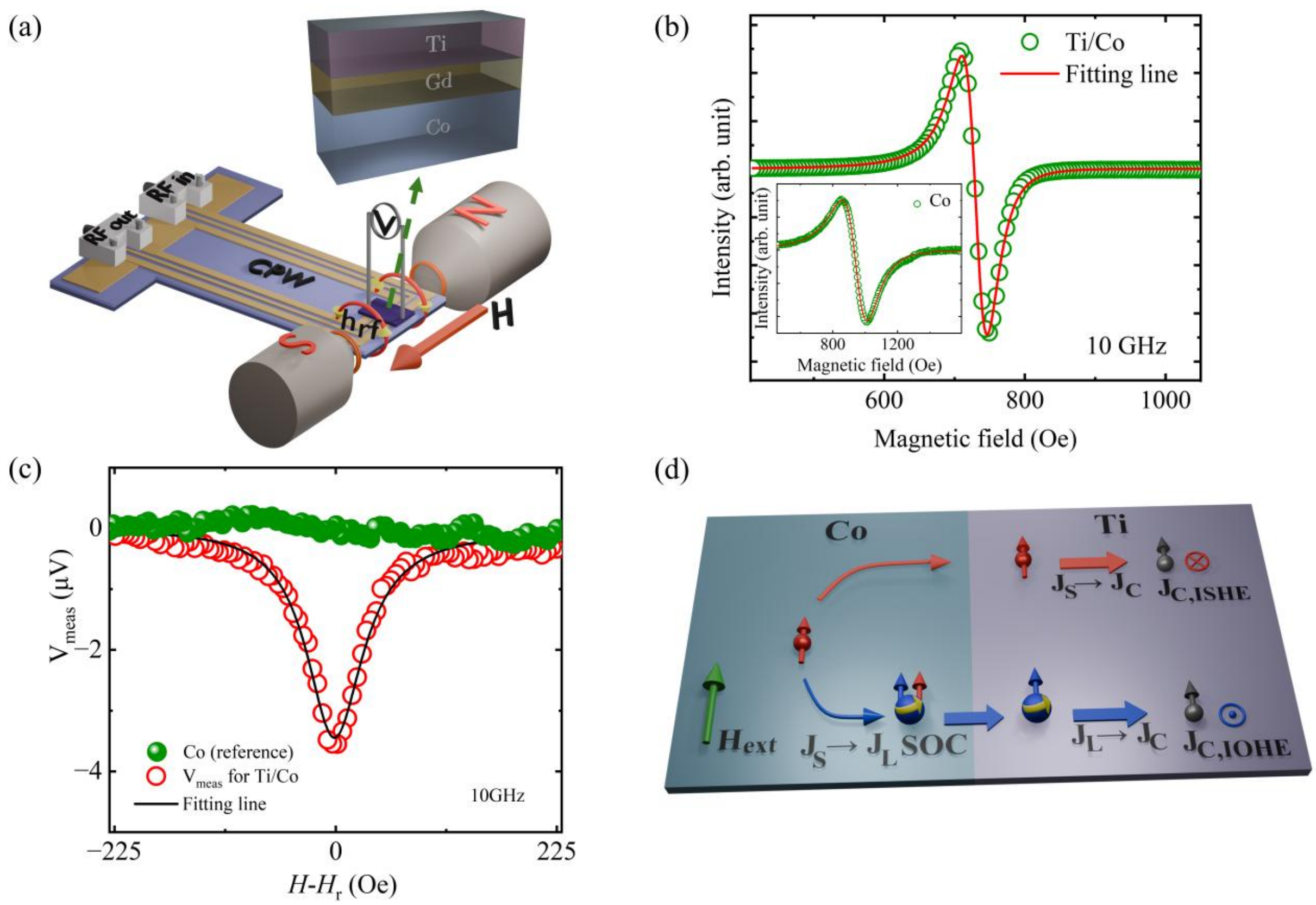

**Figure 1:** (a) Schematic of CPW-based FMR measurement set-up with transverse voltage drop detection attachment and the sample configuration (pointed by green arrow). (b) FMR absorption spectra recorded at 10 GHz for the samples Ti/Co. The inset shows FMR absorption spectra for Co. (c) The measured voltage drop across the samples Co and Ti/Co is plotted with their normalized resonance condition ($H$-$H_r$). (d) The schematic shows the different conduction channels for charge, spin and orbital angular momenta in the Co/Ti bilayer.

To investigate whether Ti can generate an orbital current and to determine whether bulk or interfacial mechanisms dominate its origin, we fabricated two stacks: a single-layer reference Co(10 nm) and a Ti(20 nm)/Co(10 nm) bilayer. Ti is a light 3$d$ transition metal with a small spin Hall angle [35]. However, its $d$-electron bands exhibit pronounced momentum-space orbital texture, giving rise to a large positive OHE and making Ti an ideal bulk orbital source [28]. The FMR absorption spectrum for Ti/Co is shown in Fig. 1(b), and the inset of Fig. 1(b) shows the FMR spectrum of Co. The FMR spectra yield resonance fields 951.5 ± 0.3 Oe for Co and 733.0 ± 0.2 Oe for Ti/Co samples. FMR spectra of Co and Ti/Co (Fig. 1 (b)), respectively, defining the detection windows (scaled as ($H$-$H_r$) in Fig. 1 (c)) for spin-orbit pumping based transverse voltage drop measurements. The reference sample Co exhibits no DC voltage at resonance, as shown in Fig. 1(c). This indicates that self-induced spin rectification, the anomalous Hall effect (AHE) in cobalt, and any inverse orbital Rashba-Edelstein effect (IOREE) at the Co/Ti(cap) interface are all negligible under the current experimental conditions. The Co sample, therefore, establishes a clean reference baseline, isolating the transverse voltage observed in any other sample. In contrast, Ti/Co exhibits a sizable voltage drop at $H_r$ [Figure 1(c)], which we attribute primarily to orbital-to-charge conversion in the bulk of the Ti layer. The *rf*-driven magnetisation precession in Co at FMR injects both a spin current $J_S$, through the finite spin-orbit correlation $\langle L \cdot S \rangle_{Co}$ of the ferromagnet, and an orbital current $J_L$ into the adjacent Ti layer, as schematically presented in Fig. 1(d). The orbital pumping efficiency is proportional to $\langle L \cdot S \rangle_{Co}$, which is non-negligible for Co. Because the spin Hall angle of Ti ($|\theta_{SH}^{Ti}|$) is negligibly small, the inverse spin Hall contribution from $J_S$ in Ti is negligible. Therefore, our results suggest that $J_L$ propagates through the Ti and is converted to a transverse charge current $J_C$ via the IOHE, driven by $\theta_{OH}^{Ti}$. An additional contribution from an interfacial IOREE at the Ti/Co boundary cannot be fully excluded in this bilayer geometry, as both the bulk IOHE and the interfacial IOREE generate charge currents of identical symmetry. To decouple these two channels, a nonmagnetic interlayer is introduced between Ti and Co, as discussed in the subsequent section.

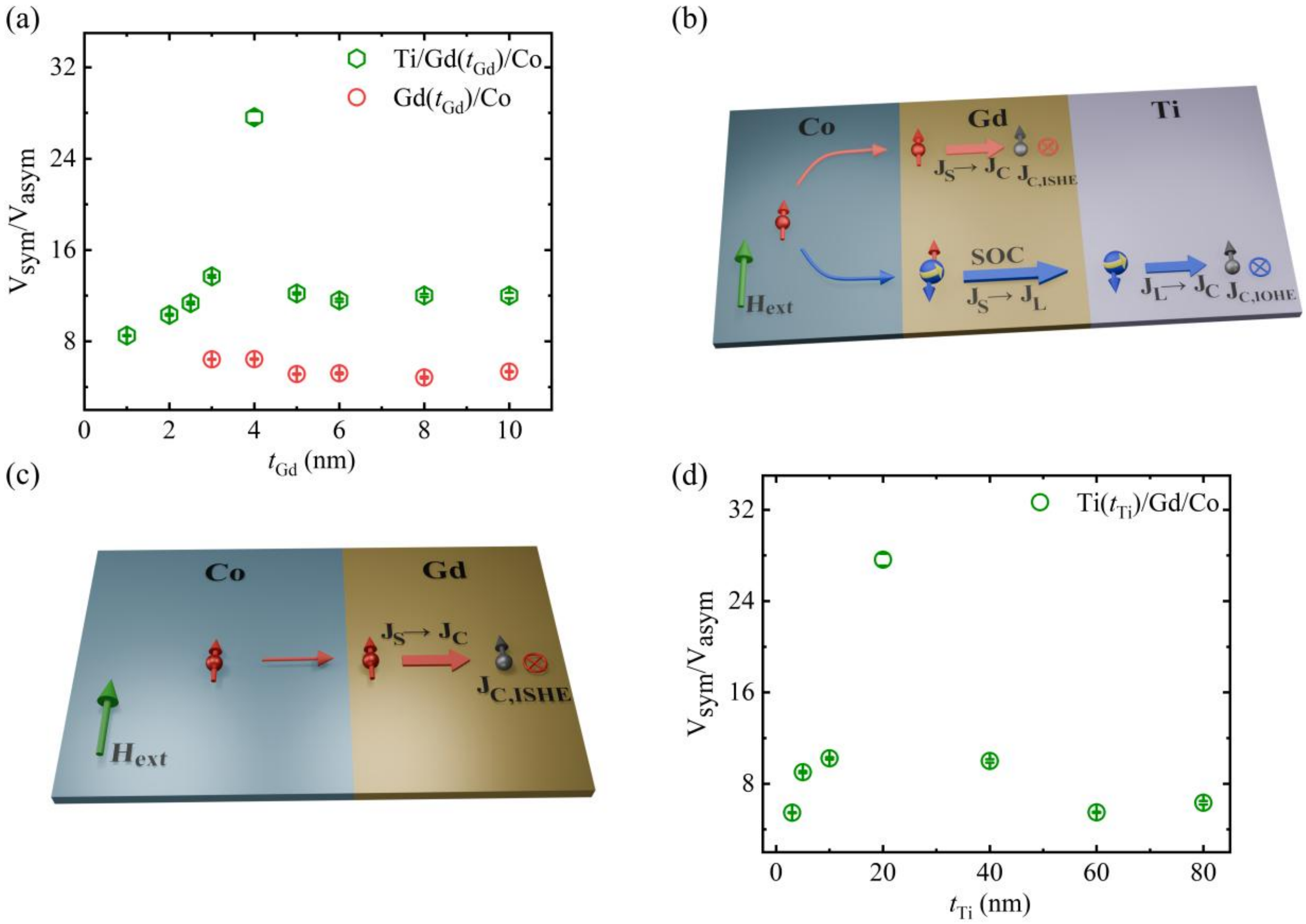

**Figure 2:** (a) $V_{sym}/V_{asym}$ ratio as a function of Gd thickness ($t_{Gd}$) for the sample set with inclusion of Ti (green hexagons) or without Ti (red circles). (b) and (c) Schematic representation of charge, spin and orbital current interconversion and conduction channels for Ti/Gd/Co trilayer and Gd/Co bilayer in spin-orbit pumping-based measurement technique. (d) $V_{sym}/V_{asym}$ ratio as a function of Ti thickness ($t_{Ti}$) in Ti($t_{Ti}$)/Gd/Co sample set.

According to Hund's third rule, the spin-orbit correlation in light elements has the opposite sign to that in ferromagnetic transition metals, which constrains the orbital-to-spin conversion in a simple LE/FM bilayer. As noted earlier, this limitation, together with interfacial orbital scattering, reduces the effective contribution of the OHE to the total torque [22,36]. To address these limitations simultaneously, we introduce a rare-earth gadolinium (Gd) IL between Ti and Co. Owing to its strong spin-orbit coupling and spin-orbit correlation $\langle L\cdot S\rangle_{Gd}$, Gd is expected to serve concurrently as (i) an interface modulator that alters the interfacial scattering environment and facilitate smooth orbital momentum at the Ti/Co boundary, (ii) a facilitator of spin-to-charge ($J_S \rightarrow J_C$) conversion via the inverse spin Hall effect, and (iii) a mediator of spin-to-orbital ($J_S \rightarrow J_L$) interconversion that relays orbital current to the adjacent Ti layer for subsequent IOHE detection. The measured DC voltage $V_{meas}$ generated at FMR is decomposed into a symmetric Lorentzian component $V_{sym}$ and an antisymmetric component $V_{asym}$ according to the standard fitting procedure [37,38].

$$V_{meas} = V_{sym} \frac{(\Delta H)^2}{(H-H_r)^2+(\Delta H)^2} + V_{asym} \frac{-2\Delta H(H-H_r)}{(H-H_r)^2+(\Delta H)^2} \quad (1)$$

$V_{sym}$ predominantly arises from the damping-like angular momentum injection associated with the spin pumping (SP) and/or orbital pumping (OP). In contrast, $V_{asym}$ has contributions from the field-like torque, the Oersted field, and spin (orbital) rectification effects, including the anisotropic magnetoresistance (AMR) and AHE of the FM layer. We, therefore, define the ratio $R(= \frac{V_{sym}}{V_{asym}})$ as a qualitative figure of merit. Larger values of $R$ indicate that the symmetric, spin-pumping or orbital-pumping-dominated contribution accounts for a greater fraction of the total voltage. It implies that spin-charge or orbital-charge conversion dominates over purely rectification-based contributions. The Gd-thickness dependence of $R$ is used in the following to identify the orbital current pathway and to distinguish SP from OP contributions in the Ti/Gd/Co heterostructure. Figure 2(a) shows the ratio, $R$, as a function of the Gd interlayer thickness ($t_{Gd}$) for Ti(20)/Gd($t_{Gd}$)/Co(10) multilayers. We observe a continuous increment in $R$ value as $t_{Gd}$ increases from 1 nm to 3 nm. In the lowest thickness limit, continuous Gd bulk has not yet formed, and only interfacial scattering processes are operative. The spin current accumulated at the Co/Gd interface cannot propagate efficiently into the Gd bulk. As $t_{Gd}$ increases, the Gd layer becomes sufficiently thick to convert the interfacial spin current more effectively. $R$ exhibits a sharp increase and attains its maximum at $t_{Gd} \approx$ 4 nm [Fig. 2 (a)], which we ascribe to the simultaneous fulfilment of two conditions. First, the Gd layer is sufficiently thick to sustain bulk spin-orbit coupling, enabling efficient $J_S \rightarrow J_C$ and $J_S \rightarrow J_L$ conversion through its intrinsic spin Hall and orbital Hall responses. Second, at this thickness, the Gd layer remains below its orbital diffusion length $\lambda_L^{Gd}$, permitting the largest fraction of $J_L$ to traverse to the Ti/Gd interface, where it is subsequently converted into a charge current via the IOHE in Ti, as schematically described in Fig. 2(b). This picture is consistent with the orbital diffusion length of Gd reported to be ~ 3 nm [36], which is close to our experimental observation at a nominal thickness of 4 nm. Therefore, the orbital diffusion length can be qualitatively considered to be ~ 4 nm for Gd, in our samples. The enhanced $R$ observed for $t_{Gd} \approx 4$ nm therefore reflects the superposition of an ISHE contribution from $J_S \rightarrow J_C$ conversion in Gd and an IOHE contribution from $J_L \rightarrow J_C$ conversion in Ti, driven by orbital current transmitted through the Gd interlayer. The dominant contribution is attributed to the IOHE. To validate this attribution, we have prepared Gd($t_{Gd}$)/Co and measured spin-pumping assisted (as Ti is not present) transverse voltage drop. These samples exhibit a lower $R$ value as compared to the Ti/Gd($t_{Gd}$)/Co, confirming that the enhanced signal in Ti/Gd($t_{Gd}$)/Co is dominated by orbital pumping and subsequent IOHE conversion in Ti. For Gd($t_{Gd}$) > 4 nm, $R$ decreases monotonically and approaches a saturated value consistent with the conventional ISHE signal generated by $J_S \rightarrow J_C$ conversion within the Gd layer itself. Beyond its orbital diffusion length, orbital current carried by Gd is absorbed and dissipated within the Gd bulk before reaching the Ti layer. Consequently, the IOHE contribution from Ti is effectively suppressed. In this thick-Gd limit, the symmetric voltage $V_{sym}$ is dominated by the inverse spin Hall mechanism in Gd (schematic representation is shown in Fig. 2 (c)), which saturates once Gd($t_{Gd}$) exceeds the $\lambda_L^{Gd}$, for set B. In the absence of Ti, $R$ increases slightly with increasing Gd thickness between 3 and 4 nm and then saturates. This saturation is consistent with a purely spin Hall-mediated ISHE signal from Gd, which is expected to reach a constant value once thickness of Gd substantially exceeds its spin diffusion length $\lambda_S^{Gd}$. Critically, the $R$ values for sample set Ti/Gd($t_{Gd}$)/Co are different and slightly higher as compared to sample set Gd($t_{Gd}$)/Co. This residual enhancement is attributed to a persistent, non-negligible orbital current $J_L$ even at Gd thicknesses moderately exceeding its orbital diffusion length. This orbital current can still diffuse from Gd into Ti and contributes to the IOHE voltage. The difference in $R$ between the Ti/Gd($t_{Gd}$)/Co and Gd($t_{Gd}$)/Co stacks further confirms that the enhanced symmetric voltage observed in the Ti-containing structures arises from the IOHE and cannot be attributed solely to the ISHE contribution of Gd.

A rigorous quantitative extraction of the effective conversion efficiency from the present data is nontrivial because several spin-orbital conversion channels contribute simultaneously. These include $J_L \rightarrow J_C$ in Ti and both $J_S \rightarrow J_C$ and $J_S \rightarrow J_L$ in Gd. Consequently, the effective efficiency depends on several parameters, including orbital diffusion lengths $\lambda_L^{Ti}$, $\lambda_L^{Gd}$and the spin diffusion length $\lambda_S^{Gd}$. The inclusion of a spacer layer that is itself an active spin-orbital converter introduces fundamental ambiguity into the standard drift-diffusion framework, as neither $\lambda_L^{Ti}$ nor $\lambda_S^{Gd}$ can be independently determined without additional assumptions. This limitation is further reinforced by the absence of consensus values for $\lambda_L^{Ti}$ in the literature. Reported values vary significantly depending on the experimental geometry, detection scheme, and angular-momentum injection mechanism. For example, Santos et al., employing an orbital pumping scheme with drift-diffusion modelling, extracted $\lambda_L^{Ti} \approx 4$ nm [39] whereas Hayashi et al. reported $\lambda_L^{Ti} \approx 4.6$ nm employing IOHE measurement technique [40]. These are an order of magnitude smaller than the value reported by Choi et al. i.e. $\lambda_L^{Ti} \approx 74 \pm 24$ nm via magneto-optical detection of the OHE in Ti [28]. The discrepancy among these values likely reflects the strong sensitivity of orbital transport to interface quality, crystalline structure, and the specific angular momentum injection mechanism, which remains an open challenge in the field. Accordingly, the present analysis should be regarded as qualitative. By tracking $R$ as a function of $t_{Gd}$ and comparing Ti/Gd/Co with Gd/Co control structures within the established ISHE detection framework, we decouple the relative contributions of the Gd spacer and the Ti detection layer. This approach allows us to assess their respective roles without assigning a definitive numerical value of $\lambda_L^{Ti}$.

To unravel the Ti/Gd interface contributions from the bulk of Ti, we further investigated Ti thickness variation while keeping the Gd and Co thicknesses fixed. We evaluated $R$ as a function of Ti($t_{Ti}$) in Ti($t_{Ti}$)/Gd(4 nm)/Co(10 nm) multilayers with the Gd interlayer fixed at its optimal thickness (Fig. 2(d)). At low $t_{Ti}$, only a fraction of the orbital current $J_L$ relayed by the Gd interlayer into Ti is converted to charge current, as the IOHE efficiency scales as $\tanh(t_{Ti}/2\lambda_L)$ in the diffusive regime. Nevertheless, $R$ already exceeds the Ti-free Gd(4 nm)/Co(10 nm) reference at $t_{Ti} = 5$ nm and increases monotonically up to a maximum at $t_{Ti} \approx 20$ nm. This continuous growth is inconsistent with a purely interfacial mechanism, which would saturate within a few nanometres, and confirms that the signal originates from the bulk orbital Hall response of Ti. Beyond 20 nm, $R$ decreases and approaches saturation in the range 40-80 nm. Within the orbital drift-diffusion model, this saturation indicates the onset of the regime $t_{Ti} > \lambda_L^{Ti}$, beyond which orbital current accumulated in the Gd/Ti interface dissipates before contributing efficiently to the IOHE voltage. We therefore obtain a qualitative estimate of $\lambda_L^{Ti} > 20$ nm in the present study, lying intermediate between the values of ≈ 4.6 nm and ≈ 61 nm reported in the literature [28,39,41]. In the thick-Ti limit ($t_{Ti} > 40$ nm), $R$ converges to the Ti-free reference value, confirming that $V_{sym}$ is then dominated by the conventional ISHE in the Gd interlayer with the IOHE contribution from Ti fully suppressed.

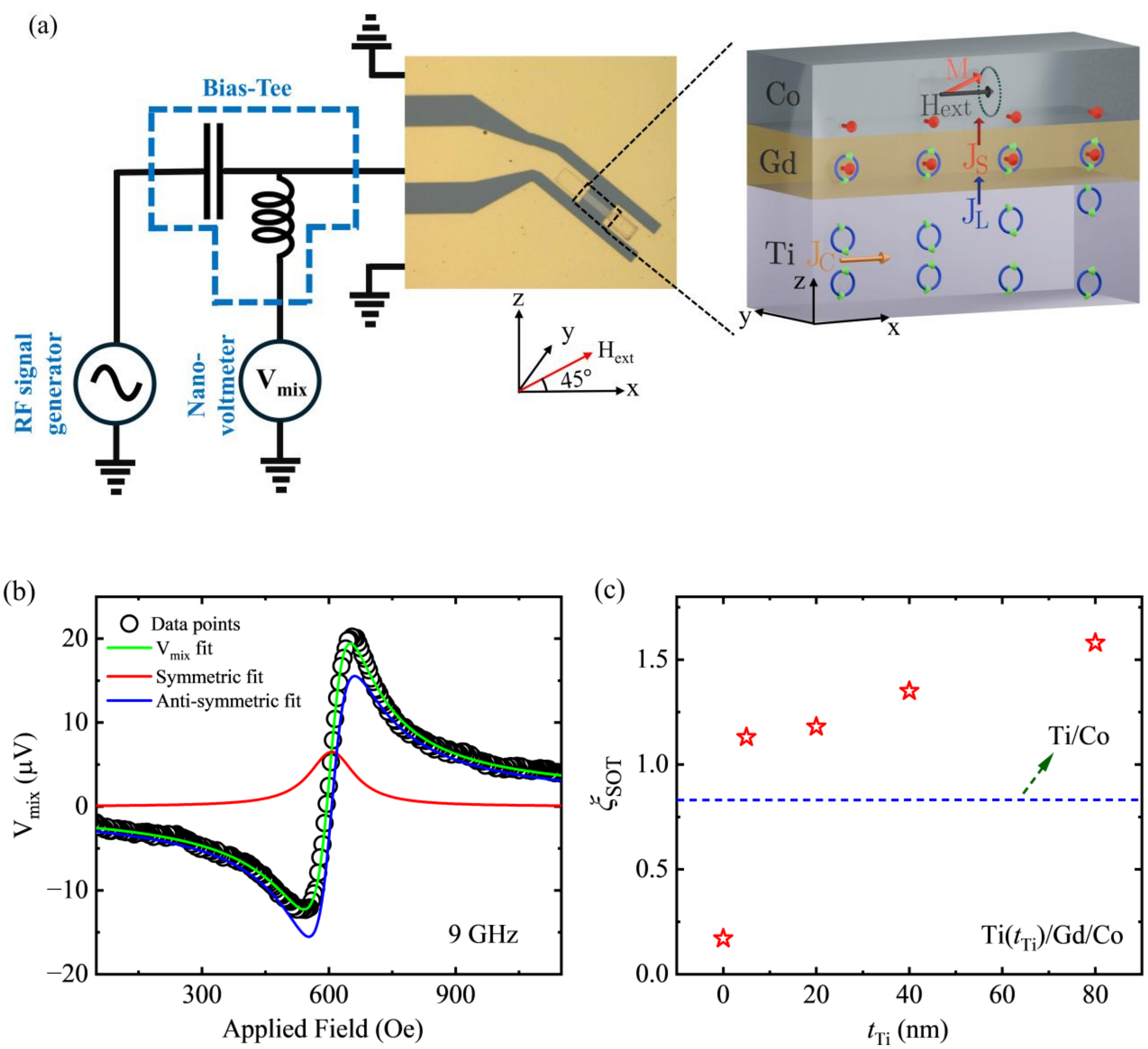

**Figure 3:** (a) An optical image of the microfabricated device incorporated with a schematic of ST-FMR measurement set up is shown. Sample geometry with torque direction is also highlighted. (b) An ST-FMR-generated mixing voltage (measured at 9 GHz) with a fitting is shown. The decomposed symmetric and anti-symmetric components with fitting lines are also presented. (c) SOT efficiency ($\xi_{SOT}$) in Ti($t_{Ti}$)/Gd/Co samples as a function of Ti thickness is shown. The horizontal dotted line represents $\xi_{SOT}$ of the control Ti/Co sample.

To place the observed experimental results on a quantitative basis, we measure the SOT efficiencies by using ST-FMR in Ti($t_{Ti}$)/Gd/Co and Ti/Co samples on lithographically patterned bar devices. An optical micrograph of the patterned device together with the measurement circuit is shown in Fig. 3(a). A radio-frequency current $I_{RF}$ is injected along the device length while a static in-plane magnetic field $H$ is swept from 0 to 1200 Oe at a fixed angle φ = 45° relative to $I_{RF}$. The oscillating spin-orbit torques generated by $I_{RF}$ excite magnetization precession in the Co layer at resonance. This produces a periodic oscillation in the device magnetoresistance. Mixing of this oscillating resistance with $I_{RF}$ yields a measurable DC voltage, $V_{mix}$ that is decomposed into symmetric ($V_{sym}$) and antisymmetric ($V_{asym}$) Lorentzian components according to [26, 32]

$$V_{mix} = V_{sym}\frac{(\Delta H)^2}{(H-H_r)^2+(\Delta H)^2} + V_{asym}\frac{\Delta H(H-H_r)}{(H-H_r)^2+(\Delta H)^2} \quad (2)$$

where $V_{sym}$ and $V_{asym}$ are the symmetric and antisymmetric Lorentzian components at the resonance frequency, and $\Delta H$ and $H_r$ denote the linewidth of the resonance signal and resonance field, respectively. Representative $V_{mix}$ spectra together with the resolved ($V_{sym}$) and ($V_{asym}$) components for a reference sample are shown in Fig. 3 (b).

The value of effective saturation magnetization $4\pi M_{eff}$ can be calculated by fitting to the Kittel equation $f = \gamma_{eff}/2\pi\sqrt{H_r(H_r + 4\pi M_{eff})}$, where $\gamma_{eff}$ represents the effective gyromagnetic ratio. Extracting $V_{sym}$ , $V_{asym}$ from equation (2) and $4\pi M_{eff}$ from the Kittel equation, we have evaluated the SOT efficiency using the equation [22,42,43]

$$\xi_{SOT} = \left(\frac{V_{sym}}{V_{asym}}\right)\left(\frac{e\mu_0 M_s d_{NM} t_{FM}}{\hbar}\right)\sqrt{1 + \left(\frac{4\pi M_{eff}}{H_r}\right)} \quad (3)$$

where $e$ is the elementary charge, $\mu_0$ refers to the vacuum permeability, $M_s$ stands for saturation magnetization value, $d_{NM}$ and $t_{FM}$ are the film thicknesses of the nonmagnetic layer and Co layers, respectively, and $\hbar(= h/2\pi)$ is the reduced Planck's constant. Figure 3(c) presents $\xi_{SOT}$ as a function of Ti thickness for the Ti(t)/Gd(4 nm)/Co(10 nm) sample series, together with a reference value for Ti(20 nm)/Co(10 nm) (dashed reference line). The Ti-free sample Gd(4)/Co(10) exhibits extremely low $\xi_{SOT}$, consistent with the small spin Hall angle of Gd and the absence of an orbital current source. In contrast, the Ti(20)/Co(10) reference shows an approximately fivefold enhancement in $\xi_{SOT}$ over the Gd(4)/Co(10) sample. This enhancement is attributed to the large positive OHE of Ti generating orbital torque on the Co magnetization through orbital-to-spin conversion facilitated by the finite spin-orbit correlation $\langle L\cdot S\rangle_{Co}$ of the ferromagnet. The absence of Gd in Ti(20)/Co(10) means that only the intrinsic L-S conversion within Co is operative; the substantially higher $\xi_{SOT}$ compared with the Gd-only reference therefore quantifies the orbital torque contribution of the Ti underlayer directly. Similar experimental results have also been observed by Bony et al, where a large damping-like torque has been observed in Co/Cu* bilayers without the need of orbital to spin conversion in the Pt layer [44]. It is noteworthy that the combined Ti(t)/Gd(4)/Co(10) structures exceed both individual reference values for Ti with different thicknesses, with the enhancement prominent even at $t_{Ti} = 5$ nm. This demonstrates that the Gd interlayer efficiently amplifies the orbital torque from Ti through enhanced orbital-to-spin conversion. The Gd layer converts the $J_L$ generated by Ti into a $J_S$ of augmented amplitude, which then exerts a damping-like torque on the Co magnetization. The resulting $\xi_{SOT}$ therefore reflects the combined effect of the OHE of Ti and the orbital-to-spin conversion efficiency of the Gd interlayer, both of which contribute constructively. Furthermore, $\xi_{SOT}$ increases monotonically with $t_{Ti}$, consistent with a bulk orbital Hall origin in Ti and ruling out a contribution confined to the Ti/Gd interface. Taken together, these results provide quantitative evidence that integrating the light-metal orbital source Ti with the spin-orbit-active rare-earth layer Gd substantially enhances the attainable SOT efficiency beyond that of either constituent alone. This synergistic heterostructure demonstrates that judicious orbital-to-spin conversion engineering can overcome the limitations of the individual materials. Our findings establish a practical materials platform for orbital-torque-based spintronic device applications.

**Conclusion:**

In summary, we have systematically demonstrated that the principal barriers limiting orbital-torque efficiency in light-metal/ferromagnet heterostructures can be mitigated through rare-earth interlayer engineering. Ferromagnetic resonance-based spin-orbit pumping measurements on Ti(20 nm)/Gd($t_{Gd}$)/Co(10 nm) multilayers reveal an optimal Gd interlayer thickness of ≈ 4 nm, at which the symmetric-to-antisymmetric voltage ratio $R$ attains its maximum, reflecting peak orbital-to-spin conversion efficiency. Systematic comparison of $R$ between Ti/Gd($t_{Gd}$)/Co and Gd($t_{Gd}$)/Co control stacks unambiguously assigns the dominant contribution of the enhanced symmetric voltage to the inverse orbital Hall effect in bulk Ti rather than to the inverse spin Hall effect in Gd alone. The Ti thickness dependent study in Ti($t_{Ti}$)/Gd/Co trilayers corroborates this interpretation: $R$ increases monotonically with Ti thickness, inconsistent with a purely interfacial mechanism, and saturates beyond 20 nm, yielding a qualitative orbital diffusion length exceeding 20 nm. ST-FMR measurements on lithographically patterned devices provide quantitative validation, the SOT efficiency of Ti(20 nm)/Co ($\xi_{SOT}$ ≈ 0.83) is fivefold larger than that of the Gd(4 nm)/Co bilayer reference ($\xi_{SOT} \approx 0.17$), establishing Ti as an efficient bulk orbital-torque source without any heavy spin-orbit-coupled material in the stack. Critically, the trilayer Ti/Gd/Co architecture surpasses both bilayer structures across the entire range of Ti thicknesses investigated, with $\xi_{SOT}$ exceeding unity. The monotonic scaling of $\xi_{SOT}$ with Ti thickness rules out an independent spin Hall contribution from the Gd spacer and confirms that Gd amplifies the Ti-generated orbital torque through a constructive orbital-to-spin conversion pathway. Taken together, these results establish rare-earth interlayer engineering as a broadly transferable design principle for overcoming the orbital-conversion bottleneck inherent to light-metal orbitronic systems. This work can potentially lead to a scalable and materials-accessible route toward next-generation, low-power spin-orbitronic devices.

**Acknowledgement:**
R.M. and A.B. thank the Ministry of Education (MoE) for the fellowship. C.R. and C.M. would like to acknowledge DRDO (ERIP/ER/202312003/M/01/1853) for financial support. A.H. thank ANRF (ANRF/ARG/2025/009623/PS) for funding through Advanced Research Grant.